\documentclass[a4paper]{article}
\usepackage{epsfig,amsfonts,amssymb,multicol,setspace,textcomp}
\usepackage[T1]{fontenc}
\usepackage{amsmath}


\renewenvironment{thebibliography}[1]{\begin{oldthebibliography}{#1}\setlength{\parskip}{0ex}\setlength{\itemsep}{0ex}}{\end{oldthebibliography}}

\begin{document}
\fontsize{11}{11} \selectfont
\title{\bf Gravitational lens equation: critical solutions\\ and magnification near folds and cusps}
\author{\textsl{A.\,N.\,Alexandrov$^{1}$, S.\,M.\,Koval$^{2}$\thanks{seregacl@gmail.com}, V.\,I.\,Zhdanov$^{1}$}}
\date{\vspace*{-6ex}}
\maketitle
\begin{center} 
{\small $^{1}$Astronomical Observatory, Taras Shevchenko National University of Kyiv, Observatorna str., 4, 04053, Kyiv,  Ukraine\\
$^{2}$National University of Kyiv-Mohyla Academy, Skovorody str., 2, 04655, Kyiv, Ukraine \\}
\end{center}
\vspace*{-4ex}
\begin{abstract}
We study approximate solutions of the gravitational lens equation and corresponding lens magnification factor near the critical point. This consideration is based on the Taylor expansion of the lens potential in powers of coordinates and an introduction of a proximity parameter characterising the closeness of a point source to the caustic. Second-order corrections to known approximate solutions and magnification are found in case of a general fold point. The first-order corrections near a general cusp are found as well.\\[1ex]
\end{abstract}
\begin{multicols}{2}

\section*{\sc introduction}
\vspace*{-1ex}
\indent\indent Main equation of gravitational lens theory (eq. (\ref{eq1}) below) sets a relation between the angular position ${\rm {\bf y}}$
of the point source and the observable  position ${\rm {\bf x}}$ of its image \cite{schneider92}. The main interest is  related to critical 
points of the two-dimensional lens mapping, i.\,e. the values of ${\rm {\bf x}}_{cr} $ where Jacobian of the lens mapping vanishes: $J\left({{\vec{\mathrm{x}}}_{cr}}\right)=\left.{{D\left({y_1,y_2}\right)}\mathord{\left/{\vphantom{{D\left({y_1,y_2}\right)}{D\left({x_1,x_2}\right)}}}\right.\kern-\nulldelimiterspace}{D\left({x_1,x_2}\right)}}\right|_{{\vec{\mathrm{x}}}_{cr}}=0$. In its turn, the image of set of critical points is a set of caustic curves. Each caustic typically appears as a closed smooth curve with the so-called cusps at some isolated points. Regular points of a  caustic as singularities of differential mapping are called folds. When a point source crosses the fold caustic, the two critical images either emerge or disappear. The critical images of the point source approach the critical curve and their brightness tends to infinity when the source comes close to fold. In the vicinity of a cusp we have three critical images with infinite brightness, but only two images disappear after crossing the 
cusp. The details can be found in \cite{schneider92}.  

The singular properties of caustic points play a key role in explanation of some qualitative features and quantitative characteristics of the gravitational lensing phenomenon. For example, the qualitative picture of quadruple lensing can be modelled by a singular isothermal ellipsoid \cite{congdon08}. Specifically, relative positions of four images and their brightness depend on position of the source with respect to the caustic \cite{keeton03,keeton05}. Another example is related to the so-called strong microlensing events, which are interpreted as a crossing of a microcaustic by an extended source. In this case  astronomical observations give us a chance to get some information about size of the source and distribution of the brightness on its surface \cite{dominik05,griger88,mineshige99}. 

The well known approximate solutions of the lens equation and expressions for magnification of each image  obtained in the lowest approximation \cite{gaudi02a, gaudi02b,schneider92} have a sense of asymptotic relations, which are performed the better, the closer the source is located relative to the caustic.
 
In the case of a fold caustic, the first-order corrections for approximate coordinates of the critical images were found in \cite{alexandrov03}; corrections for the magnification of separate images were obtained in paper \cite{keeton05}. Note that it is possible to observe only a total brightness of all microimages during a strong microlensing event. In this case, the first-order corrections for magnification of two critical images are mutually cancelled. The second-order corrections for image coordinates, as well as for the magnification, were found in papers \cite{alexandrov11, alexandrov10}. Besides, it was demonstrated on example of the strong microlensing event in image C of gravitational lens system Q2237+0305 that the second-order corrections can be statistically significant. It was made under the simplifying assumption that there is no continuous matter near the line of sight \cite{alexandrov11, alexandrov10}. Because of importance of  accounting dark matter, we generalize expressions for the second-
order corrections near the fold in the present paper.

Concerning the cusp caustic, the first-order corrections were considered in \cite{congdon08}, but some expressions in that paper require revisions. Moreover, calculations are missing logical conclusions, they were left on some intermediate stage. Therefore, the second part of our paper is dedicated to looking for complete and more compendious expressions in the first-order approximation for the coordinates of images and magnification near the cusp.

\section*{\sc lens\,equations\,near\,critical\,point}
\vspace*{-1ex}
\indent\indent The normalized lens equation has the form:
\begin{equation}\label{eq1}
\vec{\mathrm{y}} = \vec{\mathrm{x}}-\vec{\nabla}\Phi\left(\vec{\mathrm{x}}\right),
\end{equation}
\noindent
where $\Phi\left(\vec{\mathrm{x}}\right)$ is the lens potential. This equation relates every point $\vec{\mathrm{x}}=\left(x_{1},x_{2}\right)$ of the image plane to the point $\vec{\mathrm{y}}=\left(y_{1},y_{2}\right)$ of the source plane. In the general case, there are several solutions $\vec{\mathrm{X}}_{\left(l\right)}\left(\vec{\mathrm{y}}\right)$ of the lens equation \eqref{eq1} that represent images of a point source at $\vec{\mathrm{y}}$; we denote the solution number by the index in parentheses.

Potential  $\Phi\left(\vec{\mathrm{x}}\right)$ satisfies equation $\Delta\Phi=2k$, where $k\left(\vec{\mathrm{x}}\right)$ is the density of continuous matter on the line of sight normalized on the so-called critical density. The magnification factor of each separate image is $K_{(l)} \left(\vec{\mathrm{y}}\right)=1\mathord{\left/{\vphantom{1{\left|{J\left({\vec{\mathrm{X}}_{\left(l\right)}\left(\vec{\mathrm{y}}\right)}\right)}\right|}}}\right.\kern-\nulldelimiterspace}{\left|{J\left({\vec{\mathrm{X}}_{\left(l\right)}\left(\vec{\mathrm{y}}\right)}\right)}\right|}$, where $J\left(\vec{\mathrm{x}}\right)\equiv\left|{{D\left(\vec{\mathrm{y}}\right)}\mathord{\left/{\vphantom{{D\left(\vec{\mathrm{y}}\right)}{D\left(\vec{\mathrm{x}}\right)}}}\right.\kern-\nulldelimiterspace}{D\left(\vec{\mathrm{x}}\right)}}\right|$ is the Jacobian of the lens mapping \eqref{eq1}. 

Recall that, critical curves of mapping \eqref{eq1} are determined with equation $J\left(\vec{\mathrm{x}}\right)=0$. Caustics are images of  critical curves obtained with mapping \eqref{eq1}. The stable critical points of a two-dimensional mapping can be folds and cusps only.

Using standard approach to examine neighbourhood of the caustic, potential near the point $p_{cr}$ of the critical curve can be approximated with the Taylor polynomial. Let this point be the coordinate origin. We suppose that eq.\,\eqref{eq1} maps $p_{cr}$ onto the coordinate origin of the source plane. Then, we rotate synchronously the coordinate systems until the abscissa axis on the source plane becomes tangent to the caustic at the origin; the quantity $\left|{y_{2}}\right|$ defines locally the distance to the caustic and $y_{1}$ is a displacement along the tangent. 

With a sufficient accuracy, the lens equations have the following form:
\begin{multline}\label{eq2a}
 y_{1} = 2\left({1-k_{0}}\right)x_{1}+a_{1}x_{1}^{2}-a_{2}x_{2}^{2}+2b_{2}x_{1}x_{2}+ \\ 
 +c_{2}x_{1}^{3}-3c_{1}x_{1}x_{2}^{2}-d_{1}x_{2}^{3}+3d_{2}x_{1}^{2}x_{2}+g_{1}x_{2}^{4}+... \\ 
 y_{2}=b_{2}x_{1}^{2}-b_{1}x_{2}^{2}-2a_{2}x_{1}x_{2}+d_{2}x_{1}^{3}- \\ 
 -3d_{1}x_{1}x_{2}^{2}+c_{2}x_{2}^{3}-3c_{1}x_{2}x_{1}^{2}+f_{3}x_{2}^{4}+... 
\end{multline}
Here $k_{0}=k\left(0\right)$ is the matter density at the origin and the following notations are:
\begin{multline*}
a_{1}=-{\Phi,_{111}}\mathord{\left/{\vphantom{{\Phi,_{111}}2}}\right.\kern-\nulldelimiterspace}2;
a_{2}={\Phi,_{122}}\mathord{\left/{\vphantom{{\Phi,_{122}}2}}\right.\kern-\nulldelimiterspace}2;
b_{1}={\Phi,_{222}}\mathord{\left/{\vphantom{{\Phi,_{222}}2}}\right.\kern-\nulldelimiterspace}2;\\
b_{2}=-{\Phi,_{112}}\mathord{\left/{\vphantom{{\Phi,_{112}}2}}\right.\kern-\nulldelimiterspace}2;
c_{1}={\Phi,_{1122}}\mathord{\left/{\vphantom{{\Phi,_{1122}}6}}\right.\kern-\nulldelimiterspace}6;
c_{2}=-{\Phi,_{2222}}\mathord{\left/{\vphantom{{\Phi,_{2222}}6}}\right.\kern-\nulldelimiterspace}6;\\
d_{1}={\Phi,_{1222}}\mathord{\left/{\vphantom{{\Phi,_{1222}}6}}\right.\kern-\nulldelimiterspace}6;
d_{2}=-{\Phi,_{1112}}\mathord{\left/{\vphantom{{\Phi,_{1112}}6}}\right.\kern-\nulldelimiterspace}6;
g=-{\Phi,_{12222}}\mathord{\left/{\vphantom{{\Phi,_{12222}}{24}}}\right.\kern-\nulldelimiterspace}{24};\\
f=-{\Phi,_{22222}}\mathord{\left/{\vphantom{{\Phi,_{22222}}{24}}}\right.\kern-\nulldelimiterspace}{24}.
\end{multline*}
When density $k$ is constant, then $a_{1}=a_{2}=a$, $b_{1}=b_{2}=b$, $c_{1}=c_{2}=c$, $d_{1}=d_{2}=d$. Parameter $d_{2} $ will not appear in the following formulae; therefore we put $d_{1}=d$.

\vspace*{-3ex}
\section*{\sc approximate formulae\\ near fold caustic}
\vspace*{-1ex}
\indent\indent One of the approaches for finding critical solutions of eq.\,\eqref{eq1}  involves an expansion of the image coordinates into series in powers of some parameter $t$, which demonstrates proximity to the caustic \cite{alexandrov11}-\cite{alexandrov10}. If we put $y_{i}=t^{2}\tilde{y}_{i}$, then, as it was shown in \cite{alexandrov11}-\cite{alexandrov10}, the critical solutions of eq.\,\eqref{eq1} 
are analytical functions of parameter $t$, and $x_{1}=t^{2}\tilde{x}_{1}$, $x_{2}=t\tilde{x}_{2}$, where $\tilde{x}_{1}\left(t\right)$, $\tilde{x}_{2}\left(t\right)$ are zero-order functions. Putting these expressions into Taylor expansion of eq.\,\eqref{eq1}, and restricting our solutions to second-order terms inclusive, we get the following equations:
\begin{multline}\label{eq2}
\tilde{y}_{1}=2\left({1-k_{0}}\right)\tilde{x}_{1}-a_{2}\tilde{x}_{2}^{2}+t\left({2b_{2}\tilde{x}_{1}\tilde{x}_{2}-d\tilde{x}_{2}^{3}}
\right)+\\ 
+t^{2}\left({a_{1}\tilde{x}_{1}^{2}-3c_{1}\tilde{x}_{1}\tilde{x}_{2}^{2}+g\tilde{x}_{2}^{4}}\right), \\ 
\tilde{y}_{2}=-b_{1}\tilde{x}_{2}^{2}+t\left({-2a_{2}\tilde{x}_{1}\tilde{x}_{2}+c_{2}\tilde{x}_{2}^{3}}\right)+\\ 
+t^{2}\left({b_{2}\tilde{x}_{1}^{2}-3d\tilde{x}_{1}\tilde{x}_{2}^{2}+f\tilde{x}_{2}^{4}}\right).
\end{multline}
After performing calculations, it is enough to put $t=1$ and thus return to the initial variables $y_{i}$ and $x_{i}$.

A condition that initial point $p_{cr}$ is a fold is that $b_{1}\ne0$. Without losing generality of our approach, we assume that $b_{1}<0$. When density $k$ is constant, then $a_{1}=a_{2}=a$, $b_{1}=b_{2}=b$, $c_{1}=c_{2}=c$. Therefore, the system \eqref{eq2} includes four additional parameters in comparison with previous case of \cite{alexandrov11}-\cite{alexandrov10} where $k\left(\vec{\mathrm{x}}\right)\equiv0$. 

We seek solutions of equations \eqref{eq2} accurate within second-order terms in a form: $\tilde{x}_{1}=x_{10}+x_{11}t+x_{12}t^{2},\mbox{ }\tilde{x}_{2}=x_{20}+x_{21}t+x_{22}t^{2}.$ Imposing notations $R^{2}=a_{2}^{2}+b_{1}b_{2}$, $\sigma=1-k_{0}$ and $\varepsilon=\pm1$ we find  the following expressions in the zero-order approximation:
\begin{equation}\label{eq3}
x_{10}=\frac{1}{2\sigma}\left({\tilde{y}_{1}-{a_{2}\tilde{y}_{2}}\mathord{\left/{\vphantom{{a_{2}\tilde{y}_{2}}{b_{1}}}}\right.\kern-\nulldelimiterspace}{b_{1}}}\right), \quad
x_{20}=\varepsilon\sqrt{{\tilde{y}_{2}}\mathord{\left/{\vphantom{{\tilde{y}_{2}}{\left|{b_{1}}\right|}}}\right.\kern-\nulldelimiterspace}{\left|{b_{1}}\right|}}.
\end{equation}
Two signs of parameter $\varepsilon $ correspond to two critical solutions. The first approximation gives:
\begin{multline}\label{eq4}
x_{11}=-\displaystyle{\frac{\varepsilon}{2b_{1}^{2}\sigma^{2}}}\sqrt{{\tilde{y}_{2}}\mathord{\left/{\vphantom{{\tilde{y}_{2}}{\left|{b_{1}}\right|}}}\right.\kern-\nulldelimiterspace}{\left|{b_{1}}\right|}}\left\{{b_{1}R^{2}\tilde{y}_{1}-}\right.\\ 
\left.{-\left[{a_{2}R^{2}-\left({b_{1}d+a_{2}c_{2}}\right)\sigma}\right]\tilde{y}_{2}}\right\}, 
\end{multline}
\begin{equation}\label{eq5}
x_{21}=\frac{-a_{2}b_{1}\tilde{y}_{1}+\left({a_{2}^{2}-c_{2}\sigma}\right)\tilde{y}_{2}}{2b_{1}^{2}\sigma}.
\end{equation}
Concerning the second-order approximation for the first coordinate we found: 
\begin{equation}\label{eq6}
x_{12}=\frac{M_{1}\tilde{y}_{1}^{2}+M_{2}\tilde{y}_{1}\tilde{y}_{2}-M_{3}\tilde{y}_{2}^{2}}{8b_{1}^{4}\sigma^{3}},
\end{equation}
where
\begin{equation}\label{eq7}
M_{1}=b_{1}^{2}\left({3a_{2}b_{1}b_{2}+2a_{2}^{3}-a_{1}b_{1}^{2}}\right),
\end{equation}
\begin{multline}\label{eq8}
 M_{2}=2b_{1}\left[b_{1}^{2}\left({a_{1}a_{2}-2b_{2}^{2}-3c_{1}\sigma}\right)-\right.b_{1}\left(7a_{2}^{2}b_{2}-\right. \\ 
 \left.\left.-(b_{2}c_{2}+6a_{2}d)\sigma\right)-4a_{2}^{2}\left({a_{2}^{2}-c_{2}\sigma}\right)\right], 
\end{multline}
\begin{multline}\label{eq9}
 M_{3}=b_{1}^{2}\left[{a_{1}a_{2}^{2}-4a_{2}b_{2}^{2}+(4b_{2}d-6a_{2}c_{1})\sigma+4g\sigma^{2}}\right]+ \\ 
 +b_{1}\left[{-11a_{2}^{3}b_{2}+(16a_{2}^{2}d+6a_{2}b_{2}c_{2})\sigma-\left({4a_{2}f}\right.}\right.+ \\ 
 \left.{\left.{+6c_{2}d}\right)\sigma^{2}}\right]-6a_{2}\left({a_{2}^{2}-c_{2}\sigma}\right)^{2}. 
\end{multline}
And for the second coordinate: 
\begin{equation}\label{eq10}
x_{22}=\varepsilon\sqrt{{\tilde{y}_{2}}\mathord{\left/{\vphantom{{\tilde{y}_{2}}{\left|{b_{1}}\right|}}}\right.\kern-\nulldelimiterspace} {\left|{b_{1}}\right|}}\frac{N_{1}\tilde{y}_{2}+N_{2}\tilde{y}_{1}+N_{3}{\tilde{y}_{1}^{2}}\mathord{\left/{\vphantom{{\tilde{y}_{1}^{2}}{\tilde{y}_{2}}}}\right.\kern-\nulldelimiterspace}{\tilde{y}_{2}}}{8b_{1}^{3}\sigma^2},
\end{equation}
\begin{multline}\label{eq11}
N_{1}=-5a_{2}^{2}R^{2}+10\left({a_{2}b_{1}d+a_{2}^{2}c_{2}}\right)\sigma-\\
-\left({5c_{2}^{2}+4fb_{1}}\right)\sigma^{2},
\end{multline}
\begin{equation}\label{eq12}
N_{2}=6b_{1}\left[{a_{2}R^{2}-\left({b_{1}d+a_{2}c_{2}}\right)\sigma}\right],
\end{equation}
\begin{equation}\label{eq13}
N_{3}=-b_{1}^{2}R^{2}.
\end{equation}
In its turn, for the Jacobian of the lens mapping, calculated in points where images are situated, we found:
\begin{equation}\label{eq14}
J=tJ_{0}+t^{2}J_{1}+t^{3}J_{2},
\end{equation}
\begin{equation}\label{eq15}
J_{0}=4\varepsilon\sigma\sqrt{\left|{b_{1}}\right|\tilde{y}_{2}}, \quad J_{1}=4\frac{R^{2}-c_{2}\sigma}{b_{1}}\tilde{y}_{2},
\end{equation}
\begin{equation}\label{eq16}
J{ }_{2}=\varepsilon \sqrt {{\tilde {y}_2 } \mathord{\left/ {\vphantom {{\tilde {y}_2 } {\left| {b_1 } \right|}}} \right. \kern-\nulldelimiterspace}{\left|{b_{1}}\right|}}\frac{S_{1}\tilde{y}_{2}+S_{2}\tilde{y}_{1}-N_{3}{\tilde{y}_{1}^{2}}\mathord{\left/{\vphantom {{\tilde{y}_{1}^{2}}{\tilde{y}_{2}}}}\right.\kern-\nulldelimiterspace}{\tilde{y}_{2}}}{2b_{1}^{2}\sigma},
\end{equation}
\begin{multline}\label{eq17}
 S_{1}=-11a_{2}^{2}b_{1}b_{2}+4a_{1}a_{2}b_{1}^{2}+30a_{2}b_{1}d\sigma- \\ 
 -7\left({a_{2}^{2}-c_{2}\sigma}\right)^{2}-4b_{1}\left({3b_{1}c_{1}+b_{2}c_{2}+3f\sigma}\right)\sigma, 
\end{multline}
\begin{multline}\label{eq18}
S_{2}=2b_{1}\left[{3a_{2}^{3}+5a_{2}b_{1}b_{2}-2a_{1}b_{1}^{2}}\right.\\
\left.-3\left({a_{2}c_{2}+b_{1}d}\right)\sigma\right].
\end{multline}
Take notice that formula for $J_{1}$ was found in \cite{keeton05}. Finally, for the total magnification factor of two critical images, we obtained:
\begin{equation}\label{eq19}
K_{cr}=\frac{1}{2}\frac{\Theta\left({y_{2}}\right)}{\sigma\sqrt{\left|{b_{1}}\right|y_{2}}}\left[{1+Py_{2}+Qy_{1}-\frac{\kappa}{4}\frac{y_{1}^{2}}{y_{2}}}\right],
\end{equation}
\begin{equation}\label{eq20}
P=2\kappa{b_{2}}\mathord{\left/{\vphantom{{b_{2}}{b_{1}}}}\right.\kern-\nulldelimiterspace}{b_{1}-T\mathord{\left/{\vphantom{T{8b_{1}^3\sigma^{2}}}}\right.\kern-\nulldelimiterspace}{8b_{1}^{3}\sigma^{2}}},
\end{equation}
\begin{multline}\label{eq21}
T=b_{1}\left[{19a_{2}^{2}b_{2}-4a_{1}a_{2}b_{1}-\left({30a_{2}d+12b_{2}c_{2}}\right.}\right.- \\ 
\left.{\left.{-12b_{1}c_{1}}\right)\sigma+12f\sigma^{2}}\right]+15\left({a_{2}^{2}-c_{2}\sigma}\right)^{2}, 
\end{multline}
\begin{equation}\label{eq22}
Q=\frac{3a_{2}^{3}-2a_{1}b_{1}^{2}+5a_{2}b_{1}b_{2}-3\left({a_{2}c_{2}+b_{1}d}\right)\sigma}{4b_{1}^{2}\sigma^{2}},
\end{equation}
\begin{equation}\label{eq23}
\kappa=\frac{R^{2}}{2\left|{b_{1}}\right|\sigma^{2}}.
\end{equation}
In comparison with the formulae that were found under assumption of $k=0$, we shown that all functional dependencies on the coordinates $y_{i}$ remain the same. Only expressions of coefficients in terms of derivatives of potential have changed. 

\vspace*{-3ex}
\section*{\sc first\,approximation\,near\,cusp}
\vspace*{-1ex}
\indent \indent We assume that the origin of coordinates in eq.\,\eqref{eq2a} is a cusp: $b_{1}=0$. In this case, parameter of proximity is introduced by the following relations: $y_{1}=t^{2}\tilde{y}_{1}$, $y_{2}=t^{3}\tilde{y}_{2}$, $x_{1}=t^{2}\tilde{x}_{1}$, $x_{2}=t\tilde{x}_{2}$. It can be shown that coordinates of image $\tilde{x}_{i}$ (with parameterization proposed above) are analytical functions of $t$. To return to initial coordinates, it is enough to put $t=1$. We can find from formulae\,\eqref{eq2a}, accurate within first order terms, that the lens equations near cusp caustic are
\begin{multline}\label{eq24}
 \tilde{y}_{1}=2\sigma\tilde{x}_{1}-a\tilde{x}_{2}^{2}+\left({2b\tilde{x}_{1}\tilde{x}_{2}-d\tilde{x}_{2}^{3}}\right)\cdot{t},\\ 
 \tilde{y}_{2}=-2a\tilde{x}_{1}\tilde{x}_{2}+c\tilde{x}_{2}^{3}+\left({b\tilde{x}_{1}^{2}-3d\tilde{x}_{1}\tilde{x}_{2}^{2}+f\tilde{x}_{2}^{4}}\right)\cdot{t}, 
\end{multline}
where $a=a_{2}$, $b=b_{2}$, $c=c_{2}$.

We looked for solutions in the form: $\tilde{x}_{1}=x_{10}+tx_{11}$, $\tilde{x}_{2}=x_{20}+tx_{21}$. The basis for solutions construction is a cubic equation for $x_{20}$:
\begin{equation}\label{eq25}
Cx_{20}^{3}-a\tilde{y}_{1}x_{20}-\sigma\tilde{y}_{2}=0,
\end{equation}
where $C=c\sigma-a^{2}$. 

Equation \eqref{eq25} has one or three real roots depending on the sign of expression $Q=\displaystyle{\frac{\tilde{y}_{2}^{2}\sigma^{2}}{4C^{2}}-\frac{a^{3}\tilde{y}_{1}^{3}}{27C^{3}}}$, one real root when $Q>0$ and three real roots when $Q\le0$. And explicit expressions for solutions of eq.\,\eqref{eq25} are given with Cardano formulae. 

For the first coordinate in zero order approximation, we found:
\begin{equation}\label{eq26}
x_{10}=\frac{1}{2\sigma}\left({\tilde{y}_{1}+ax_{20}^{2}}\right).
\end{equation}
We do not present intermediate formulae for the first order corrections in form that repeats results of \cite{congdon08}. Instead, we give final and simplified expressions at once, which can be checked using substitution into eq.\,\eqref{eq24}. Hence, we have: 
\begin{equation}\label{eq27}
x_{21}=\frac{B_{1}\tilde{y}_{1}x_{20}^{2}+B_{2}\tilde{y}_{2}x_{20}+Cb\tilde{y}_{1}^{2}}{4\sigma{CE}},
\end{equation}
\begin{equation}\label{eq28}
x_{11}=\frac{CB_{1}\tilde{y}_{2}x_{20}^{2}+A_{1}\tilde{y}_{1}^{2}x_{20}+A_{2}\tilde{y}_{1}\tilde{y}_{2}}{4\sigma{C}^{2}E}.
\end{equation}
Here the following notations are imposed:
\begin{equation}\label{eq29}
E=a\tilde{y}_{1}-3Cx_{20}^{2}, 
\end{equation}
\begin{equation}\label{eq30}
B_{1}=6\sigma{}abc-a^{3}b-4\sigma{}a^{2}d-6\sigma^{2}cd+4\sigma{}^{2}af,
\end{equation}
\begin{equation}\label{eq31}
B_{2}=\sigma\left({5a^{2}b-10\sigma{}ad+4\sigma^{2}f}\right),
\end{equation}
\begin{equation}\label{eq32}
A_{1}=\sigma{}a\left({5bc^{2}-10acd+4a^{2}f}\right),
\end{equation}
\begin{equation}\label{eq33}
A_{2}=a^{4}b-2\sigma{}a^{2}bc+\sigma^{2}\left({6bc^{2}-10acd+4a^{2}f}\right).
\end{equation}

For Jacobian components $\tilde{J}=t^{2}\left({J_{0}+tJ_{1}}\right)$ we found the following expressions:
\begin{equation}\label{eq34}
J_{0}=-2E,
\end{equation}
\begin{equation}\label{eq35}
J_{1}=\frac{I_{1}\left({3Cx_{20}^{2}+a\tilde{y}_{1}}\right)\tilde{y}_{2}+I_{2}x_{20}\tilde{y}_{1}^{2}}{CE},
\end{equation}
where
\begin{equation}\label{eq36}
I_{1}=a^{2}b+\sigma\left({10ad-6bc}\right)-4\sigma^{2}f,
\end{equation}
\begin{equation}\label{eq37}
I_{2}=16a^{3}d-8a^{2}bc-\sigma^{2}\left({6acd-3bc^{2}+4a^{2}f}\right).
\end{equation}

The magnification factor of each image in the first approximation is given by the expression:
\begin{multline}\label{eq38}
 K=\displaystyle{\frac{1}{\displaystyle\left|J\right|}=\frac{1}{t^{2}}\frac{1}{\left|{J_{0}+tJ_{1}}\right|}}=\displaystyle{\frac{1}{t^{2}\left|{J_{0}}\right|}\left({1-t\frac{J_{1}}{J_{0}}}\right).} 
\end{multline}
While finding last equality, we took into account that $\left|{tJ_{1}/J_{0}}\right|<1$ (for small values of parameter $t$).

\vspace*{-3ex}
\section*{\sc results and conclusions}
\vspace*{-1ex}
\indent\indent The obtained formulae \eqref{eq6}-\eqref{eq18} represent expressions of the second-order corrections for image coordinates and Jacobian near fold caustic in the case of general eq.\,\eqref{eq1}. Formulae \eqref{eq19}-\eqref{eq23} describe the total magnification of two critical images in the second-order approximation with respect to proximity to the caustic. It is important to note that the functional dependence on the coordinates $y_{i}$ and on fitting parameters remain the same, as in the case of $k\left(\vec{\mathrm{x}} \right) \equiv 0$. All the differences are in expressions for the fitting parameters; these expressions have four additional constants when a continuous matter is distributed near the line of sight. The same situation will be with formulae for the magnification factor of extended sources  \cite{alexandrov11, alexandrov10} provided that we correspondingly replace coefficients $P, Q, \kappa $ and take into account that $\sigma\ne1$. Coefficients that are discussed in the 
present paper play a role of adjustable parameters in modelling observable light curves. Specifically, taking into account a continuous matter does not change anything in previous treatment of the strong magnification event in Q2237+0305 \cite{alexandrov11, alexandrov10}. Explicit dependencies of coefficients \eqref{eq20}-\eqref{eq23} on the derivatives of potential $\Phi(\vec{\mathrm{x}})$ will be important in case of modelling deflector mass distribution.

In the last section we obtained the first-order corrections for the image coordinates and the Jacobian near a cusp caustic \eqref{eq27}-\eqref{eq37}. Some inaccuracies of paper \cite{congdon08}  have been corrected, and  explicit expressions of the corrections are found in terms of the potential expansion parameters and the roots of the cubic equation\,\eqref{eq25}.

\end{multicols}
\end{document}